Large and significantly anisotropic critical current density induced by planar defects in CaKFe$_4$As$_4$ single crystals

Sunseng Pyon[1], Ayumu Takahashi[1], Ivan Veshchunov[1], Tsuyoshi Tamegai[1], Shigeyuki Ishida[2], Akira Iyo[2], Hiroshi Eisaki[2], Motoharu Imai[3], Hideki Abe[3], Taichi Terashima[3], Ataru Ichinose[4]

1 Department of Applied Physics, The University of Tokyo, 7-3-1 Hongo, Bunkyo-ku, Tokyo 113-8656, Japan
2 National Institute of Advanced Industrial Science and Technology (AIST), 1-1-1 Umezono, Tsukuba, Ibaraki 305-8568, Japan
3 National Institute for Materials Science (NIMS), Tsukuba, Ibaraki 305-0047, Japan
4 Central Research Institute of Electric Power Industry, Electric Power Engineering Research Laboratory, 2-6-1, Nagasaka, Yokosuka-shi, Kanagawa 240-0196, Japan

**Abstract**

Three independent components of critical current density, one for the *H*//*c* axis and the other two for the *H*//*ab* plane, have been studied in CaKFe$_4$As$_4$ single crystals. When the magnetic field is applied along the *c* axis, we observed fish-tail-like peaks in the *M-H* hysteresis loop, and the magnetization at higher temperatures exceeds that at lower temperatures at high fields. When the field is applied parallel to the *ab* plane, a dip structure is observed in the *M-H* hysteresis loop near the self-field. In addition, for the *H*//*ab* plane, we succeeded in separately evaluating the large and significantly anisotropic in-plane and out-of-plane $J_c$. Transmission electron microscopy revealed the presence of planar defects parallel to the *ab* plane in CaKFe$_4$As$_4$, which have not been observed in any other iron-based superconductors. We discuss the possible relationship between the anomalous $J_c$ behavior and the planar defects.

**Introduction**

The discovery of iron-based superconductors (IBSs) in 2008 [1] has prompted great interest not only in their unconventional superconducting mechanism with high transition temperatures, but also in their potential for applications. The most attractive material for applications among various kinds of IBSs is the 122-type (Ba,K)Fe$_2$As$_2$. It has a high transition temperature $T_c$ of ~38 K for bulk [2], high upper critical field, $H_{c2}$ (> 700 kOe) [3,4], and small anisotropy ($\gamma$ < 2) [4]. It has also been demonstrated that the critical current density ($J_c$) in 122-type IBS single crystals exceeds $1 \times 10^6$ A/cm$^2$ [5], and it can be enhanced further by introducing artificial defects [5–10].

Recently, a new type of IBSs, i.e., 1144-type with CaKFe$_4$As$_4$ as one of their representatives, has been found [11]. Its crystal structure is similar to 122-type compounds. CaKFe$_4$As$_4$ has a tetragonal structure (*P4/mmm*), where the Ca and K layers stack alternatively along the *c* axis [11,12]. Several properties of 1144-type compounds have been studied such as superconducting gap state [13,14], penetration depth [15,16], pressure effects on crystal structure [17,18], and so on. CaKFe$_4$As$_4$ shows similar superconducting properties to those of optimally doped (Ba,K)Fe$_2$As$_2$. $T_c$ is approximately 35 K and $H_{c2}$ is larger than 630 kOe [12]. However, alternate stacking of Ca and K along the *c* axis may lead to different physical properties. The characteristic properties of CaKFe$_4$As$_4$ caused by the unique crystal structure and common properties similar to that of 122-type compounds, such as large critical current density, should be evaluated.

In this study, phase-pure CaKFe$_4$As$_4$ single crystals are synthesized and detailed superconducting properties including anisotropic $J_c$ are evaluated. We observed a non-monotonic temperature dependence of in-plane $J_c$. When the magnetic field is applied parallel to the *ab*-plane, a pronounced dip structure in the magnetization hysteresis loop around the self-field is also observed. Furthermore, we find a very large anisotropy between the in-plane and out-of-plane $J_c$ when the field is applied parallel to the *ab*-plane. Transmission electron microscopy revealed the presence of planar defects along the *ab*-plane, which naturally explains the large anisotropy of $J_c$ when the field is applied parallel to the *ab*-plane.

**Experimental details**

Single crystals of CaKFe$_4$As$_4$ were synthesized by the self-flux method using FeAs. We used Ca pieces (99.5%), K ingots (99.5%), and FeAs powder as starting materials. FeAs was prepared by sealing stoichiometric amounts of As grains (6N) and Fe powder (99.9%) in an evacuated quartz tube and reacting them at 900 °C for 10 h after heating at 500 °C for 10 h. A mixture with a ratio of Ca : K : FeAs = 1 : 1.2 : 10 was placed in an alumina crucible in an argon-filled glove box. The alumina crucible was sealed in a tantalum tube using the arc melting method. The tantalum tube was sealed in an evacuated quartz tube. The whole assembly was heated for 5 h at 1180 °C after a preliminary heating at 650 °C for 5 h, and cooled to 1050 °C at a rate of 26 °C/h, followed by cooling to 930 °C at a rate of 1.5 °C/h for the crystal growth. Bulk magnetization was measured by a superconducting quantum interference device (SQUID) magnetometer (MPMS-5XL, Quantum Design). For the evaluation of two components of $J_c$ for the $H//ab$ plane, samples were cut into a rectangular shape using a focused ion beam. Electrical resistivity was measured with a Quantum Design physical property measurement system (PPMS) in the ac mode. For magneto-optical (MO) imaging, an iron-garnet indicator film was placed in direct contact with the sample surface and the whole assembly was attached to the cold finger of a He-flow cryostat (Microstat-HR, Oxford Instruments). MO images were acquired by using a cooled-CCD camera with 12-bit resolution (ORCA-ER, Hamamatsu). Cross-sectional observations of the single crystals were performed with a high-resolution scanning transmission electron microscopy (STEM; JEOL, JEM-2100F). The specimens to analyze using TEM were prepared by digging and milling using a focused-ion beam (FIB), which is called the microsampling technique. The final milling using FIB was conducted at an acceleration voltage of 30 kV and with a very weak ion current of approximately 10 pA without tilting the specimen. The phase identification was carried out by means of powder x-ray diffraction (XRD) with Cu-Kα radiation (Smartlab, Rigaku).

**Experimental Results**

Figure 1(a) shows the temperature dependence of magnetization at 5 Oe for a CaKFe$_4$As$_4$ single crystal. As shown in the inset of Fig. 1(a), $T_c$ defined by the onset of diamagnetism is 36.0 K and $\Delta T_c$ is less than 1 K. The superconducting transition is also confirmed by the temperature dependence of electrical resistivity as shown in Fig. 1(b). The residual resistivity ratio (RRR) defined by the ratio of resistivity just above $T_c$ and that at room temperature, $\rho(300\ K)/\rho(36\ K)$, is 15.2. This value is consistent with the former report by Meier *et al.* with $\rho(300\ K)/\rho(36\ K) \sim 15$ [12]. The residual resistivity $\rho(0\ K) = 3.8\ \mu\Omega$cm is evaluated using the data between 36 and 50 K by fitting to the formula $\rho(T) = \rho(0\ K) + AT^2$. All these data indicate that our crystals are highly pure.

Figures 2(a) and 2(b) show the in-plane electrical resistivity for the $H//c$ axis and $H//ab$ plane measured at various magnetic fields up to 90 kOe. It is clear that the superconducting transition for the $H//c$ axis shifts rapidly to lower temperatures compared to that for the $H//ab$ plane. The anisotropy parameter $\gamma = H_{c2}^{ab}/H_{c2}^{c}$ is evaluated from the temperature dependence of $H_{c2}(T)$. The temperature dependences of $H_{c2}$ are summarized in Fig. 2(c). The definitions of $T_c$ onset and $T_c$ end are shown in Fig. 2(a). Estimated $\gamma$ at $T \sim 34$ K is 2.2. This value is consistent with the former report by Meier *et al.* [12], and similar to that of 122-type compounds of (Ba,K)Fe$_2$As$_2$ with $\gamma \sim 2$-3 near $T_c$ [4].

Next we performed magneto-optical (MO) imaging of a CaKFe$_4$As$_4$ single crystal to evaluate its homogeneity and the value of in-plane $J_c$. Figure 3(c) shows an optical micrograph of the crystal, which shows a smooth surface with no visible cracks. Figures 3(a) and 3(b) display MO images of the CaKFe$_4$As$_4$ single crystal in the remanent state at (a) 5 K and (b) 35 K, respectively, after cycling the field up to 1.6 kOe for 0.2 s. At 5 K, the magnetic field does not fully penetrate the sample because of large $J_c$ and limitation of the value of the applied field. On the other hand, the magnetic field fully penetrates the sample at 35 K, and the MO image shows the critical state field profile expected for a uniform thin-plate superconductor with clear current discontinuity lines (*d*-lines). Local magnetic induction profiles at different temperatures taken along the dotted line in Fig. 3(a) are shown in Fig. 3(d). Magnetic induction profiles at higher temperatures of 30 and 35 K show roof-top patterns, indicating that the large and homogeneous current flows throughout the sample. At lower temperatures, however, the magnetic induction profiles are saturated around 700 G because of large $J_c$ and the saturation of the magnetization of the indicator film. From the value of the trapped field, $J_c$ can be roughly evaluated [19]. At 30 K, magnetic induction $\Delta B$ of 580 G is trapped. Using the approximate formula of $J_c \sim \Delta B /(t*\beta)$ with $t = 0.00048$ cm being the thickness of the sample ($\beta$ is a parameter determined by the sample dimensions and the setting of the garnet film, and $\beta = 3.64$ in the present case) [19], $J_c$ at 30 K under the self-field is evaluated as 0.33 MA/cm$^2$.

In-plane $J_c$ in a CaKFe$_4$As$_4$ single crystal was also evaluated from magnetization measurements by applying the magnetic field along the *c*-axis using the extended Bean model [5,20,21]. The inset of Fig. 4 shows the magnetic field dependence of magnetization at various temperatures. Evaluated $J_c$ as a function of temperature is summarized in Fig. 4. $J_c$ at 2 K under the self-field and 40 kOe are approximately 1.6 and 0.18 MA/cm$^2$, respectively. These values of $J_c$ are between those of Co-doped BaFe$_2$As$_2$ [7] and K-doped BaFe$_2$As$_2$ [8]. The value of $J_c$ at 30 K under the self-field is 0.29 MA/cm$^2$. This value is consistent with that evaluated from the analysis of the MO image. Figure 4 also shows the magnetic field dependence of in-plane $J_c$ at various temperatures. The magnetic field dependence of $J_c$ above 20 K becomes very weak, or $J_c$ even show a broad maximum at higher fields. Such a non-monotonic field dependence of $J_c$ is often referred to as the fish-tail effect and has been found in cuprate superconductors [22-25] and other IBSs [26-28]. In most cases, however, the evolution of the fish-tail peak is continuous and no crossings of *M-H* curves at different temperatures occur. The abrupt change of field dependence of $J_c$ around 15 K may indicate the presence of crossover in the dominant pinning mechanisms, which could be related to the specific defect structure discussed later. As a result, $J_c$ at higher temperatures (~20-25 K) becomes larger than that at lower temperatures (~5-15 K) above 20 kOe. In most IBSs, the increase of $J_c$ at higher temperatures at a fixed field has not been observed except for over-doped BaFe$_2$(As,P)$_2$ [26].

Next, we evaluated the critical current density for the *H*//*ab* plane. As shown in Fig. 5(a) and 5(b), there are two independent critical currents in this configuration, one flowing in the *ab* plane and another flowing along the *c* axis. We tentatively designate the former as $J_{c2}$ and the latter as $J_{c3}$. It is obvious that two components of $J_c$ cannot be evaluated from a single *M-H* measurement. We made two independent *M-H* measurements for the *H*//*ab* plane at each temperature and evaluated $J_{c2}$ and

$J_{c3}$. The detailed scheme to calculate the $J_{c2}$ and $J_{c3}$ is described in the Supplemental Material [29]. *M-H* loops for the two in-plane field directions are shown in Fig. 5(c) and 5(d). Both magnetic hysteresis loops show dips around zero field. These dip structures are similar to those observed in superconductors with columnar defects [7] caused by the suppression of pinning due to misalignment of vortices from the direction of columnar defects [30]. The magnetic field dependence of $J_{c2}$ and $J_{c3}$ at various temperatures is summarized in Fig. 5(e) and 5(f), respectively. Calculated $J_{c2}$ and $J_{c3}$ near zero field and higher fields corresponding to the dip and return branch of the hysteresis loop in Fig. 5(c) and 5(d) are omitted. The evaluated $J_{c2}$ and $J_{c3}$ are highly anisotropic and larger than $J_c$ when the magnetic field is applied along the *c* axis, as shown in Fig. 4. For example, $J_c$, $J_{c2}$, and $J_{c3}$ at 2 K and 15 kOe are 0.4, 5.1, and 1.0 MA/cm$^2$, respectively. $J_{c2}$ is five times larger than $J_{c3}$. This large anisotropy of $J_c$ cannot be explained by the small anisotropy of $\gamma \sim$ 2.2 determined by the resistivity measurements shown in figure 2.

First of all, let us summarize the fundamental properties of anisotropic (quasi two-dimensional) superconductors with effective mass $m_{ab}$ and $m_c$ for carrier motion along the *ab* plane and *c* axis. Let us define the anisotropy ratio $\gamma = (m_c/m_{ab})^{0.5}$. The resistivity ratio, $\rho_c/\rho_{ab}$, should be $m_c/m_{ab} = \gamma^2$. Anisotropic $H_{c2}$ for *H//c* and *H//ab* are given by $\Phi_0/2\pi\xi_{ab}^2$ and $\Phi_0/2\pi\xi_{ab}\xi_c$, respectively. So the ratio of $H_{c2}$, $H_{c2}^{ab}/H_{c2}^{c} = \gamma$. The pinning force is described as a derivatives of pinning potential $U$ with respect to characteristic length $x$, $dU/dx$. If the dimension of defects is large enough, $U$ is independent of the direction of the vortex motion for *H//ab*. In this case, the larger coherence length indicates the smaller $dU/dx$. So the pinning force for $J_{c2}$ or $J_{c3}$ is inversely proportional to the out-plane coherence length $\xi_c$ or the in-plane coherence length $\xi_{ab}$, respectively. That is why, $J_{c2}/J_{c3} = \xi_{ab}/\xi_c = \gamma$. In our crystal, estimated $\gamma = H_{c2}^{ab}/H_{c2}^{c} = \xi_{ab}/\xi_c = 2.2$. This simple consideration tells us that $J_{c2}/J_{c3}$ should be 2.2. In addition to this simple consideration, one may need to consider the path for the critical current. When the defects are extended, the superconducting current can flow only in a narrow region between defects. In the present case, the width of the planar defects is larger than the height of it, which restricts the region for $J_{c3}$ flowing parallel to the *c* axis, making $J_{c3}$ even smaller than $J_{c2}$. Namely, $J_{c2}/J_{c3}$ can be larger than $\gamma$.

To reveal the presence of anisotropic defects, observations by a high-resolution scanning transmission electron microscope (STEM) were performed. Figures 6(a) and 6(b) show STEM images on the cross section parallel and perpendicular to the *c* axis in CaKFe$_4$As$_4$, respectively. As shown in Fig. 6(a) and 6(b), planar defects are present in the whole area of the *ac* plane and *bc* plane. STEM images taken from the [100] and [010] directions are almost identical, suggesting that these defects are two-dimensional planar defects nearly parallel to the *ab*-plane. Each defect is ~50 x 10 nm$^2$, and is separated by ~50 nm and ~40 nm along the *ab*-plane and *c*-axis, respectively. These planar defects are observed as faint oval objects in the STEM image from the [001] direction, as shown in Fig. 6(c). These planar defects should produce significantly anisotropic $J_c$ for the *H//ab* plane since they suppress the motion of vortices along the $c$ axis effectively, and block the current flow along the $c$ axis. These planar defects along the $ab$ plane are not observed in any other IBSs or even in cuprate superconductors, and their origin is still unclear. It should be noted that energy-dispersive x-ray (EDX) analyses of the elemental composition show no chemical inhomogeneities around these defects. Planar defects with ~2 nm height and 20 nm width can be observed from the higher resolution STEM image taken from the [100] direction as shown in Fig. 6(d). So they are not due to simple stacking disorder of Ca and K along the $c$ axis.

Another supporting evidence for the planar defects is obtained from x-ray diffraction measurements. Figure 7 shows the single-crystal x-ray diffraction pattern of CaKFe$_4$As$_4$. Only the (00$l$) peaks are detected. Peaks with odd numbers of $l$ are the evidence for the formation of a 1144-type structure in CaKFe$_4$As$_4$ [11]. Peaks from possible impurities such as KFe$_2$As$_2$ and CaFe$_2$As$_2$ are not observed. Despite the high phase purity of the crystal, the observed peaks are broad. The inset of Fig. 7 shows a comparison between the (008) peaks for the CaKFe$_4$As$_4$ and (Ba,K)Fe$_2$As$_2$ single crystals [31]. (Ba,K)Fe$_2$As$_2$ shows a sharp single set of (008) peaks from Cu K$_{\alpha 1}$ and K$_{\alpha 2}$, while multiple (008) peaks are observed in CaKFe$_4$As$_4$. This indicates that the lattice constant along the $c$ axis is inhomogeneous. It should be caused by inhomogeneous stacking periodicity along the $c$ axis in CaKFe$_4$As$_4$ and is consistent with the presence of planar defects observed by STEM shown in Fig. 6(a).

**Discussion**

We found anomalous features of magnetic hysteresis loops in CaKFe$_4$As$_4$ single crystals. First, when the magnetic field is applied along the $c$-axis, magnetization at higher temperatures exceeds that at lower temperatures at high fields as shown in Fig. 4. One of the possible origins for that is field-induced pinning centers, which are effective only at high magnetic fields and high temperatures. Such field-induced pinning centers can be created due to the local distributions of $T_c$ in the crystal such as KFe$_2$As$_2$ and CaFe$_2$As$_2$, which are possibly embedded as impurity phases in CaKFe$_4$As$_4$ [32]. They can pin vortices only at higher temperatures than their $T_c$. Another possible origin for the anomalous temperature dependence of $J_c$ is the presence of large-scale local variation of superconducting properties due to chemical inhomogeneities. However, as shown in Fig. 3, MO images clearly show no local variation of $J_c$ in our crystal. In addition, as mentioned above, secondary phases were not detected from x-ray diffraction analyses. We infer that anomalous features of magnetic hysteresis loops are analogous to the fish-tail effect observed in other superconductors, such as cuprate [22-25] and iron-based superconductors [26-28]. For example, in the magnetization loop at 30 K, a dip at 5 kOe and broad peak at ~15 kOe are found. The origin of a fish-tail-like peak in IBSs is still not clear although it is suggested that a crossover from the collective to the plastic creep in field is the possible origin of the fish-tail peak in Ba(Fe,Co)$_2$As$_2$

[33]. A similar dip and peak in the hysteresis loop in CaKFe$_4$As$_4$ imply the same origin, although the phenomenon that magnetization at higher temperatures exceeds that at lower temperatures is very rare. Second, we have succeeded in evaluating two components of $J_c$ when the field is applied along the $ab$ plane. In this configuration, $J_{c2}$ with current along the $ab$ plane is more than five times larger than $J_{c3}$ with current along the $c$ axis. Anisotropic $J_c$ for the $H//ab$ plane has not been studied properly in other iron-based and even cuprate superconductors. For instance, Ba(Fe,Co)$_2$As$_2$ and Fe(Te,Se) single crystals are reported to show isotropic $J_c$ [7, 34]. In these reports, $J_c$ for the $H//c$ axis has a similar value to that for the $H//ab$ plane in 122 and 11 systems with the assumption that two components of $J_c$ for the $H//ab$ plane are equal. However, in CaKFe$_4$As$_4$, anisotropic $J_c$ for the $H//ab$ plane has been successfully evaluated since $J_{c2}$ is much larger than $J_{c3}$. In all studies to date, $J_{c2}$ and $J_{c3}$ are assumed to be equal and only the "average" $J_c$ for the $H//ab$ plane has been evaluated, i.e., $J_{c2} = J_{c3} = 20\Delta M/\alpha l(1-\alpha/3)$ (see Supplemental Material [29]). Third, when the magnetic field is applied along the $ab$-plane, a significant dip structure around zero field is observed. This dip structure observed in CaKFe$_4$As$_4$ has not been observed in any other pristine iron-based superconductors such as Ba(Fe,Co)$_2$As$_2$ and Fe(Te,Se) [7, 34]. Similar dip structures in the magnetization hysteresis loop around the self-field are observed in cuprate and iron-based superconductors with columnar defects when the magnetic field is applied parallel to the defects [35-37]. The dip structure might be produced by inhomogeneous magnetic field distribution around the self-field [7] because the planar defects can also enhance pinning forces similar to columnar defects. However, this mechanism for the dip can only be realized when the magnetic field is applied perpendicular to the surface of thin crystals ($H//c$ axis). When the magnetic field is applied along the large plane of the crystal, the demagnetization effect can be neglected. In such a case, the curvature of vortices due to the self-field is negligible and $J_c$ cannot be suppressed near zero field. Considering the planar defects in CaKFe$_4$As$_4$, the possibility of the matching between thee magnetic field and the defect structure can be suggested. Enhancement of $J_c$ is reported in superconducting films with a regular array of holes when the number of vortices matches with the number of holes [38,39]. The areal density of the observed defects shown in Fig. 6 is $1/(\sim 50 \times 40) = 5 \times 10^{-4}$ nm$^{-2}$, leading to the matching field of 10.3 kOe. This is very similar to the magnetic field range of the dip around the self-field shown in Fig. 5. In Fig. 8, we compared the characteristic value of fields in the hysteresis loops of $M_1$ and $M_2$ from Fig. 5(c) and 5(d). The widths of the return branch and $J_{c2}t$ or $J_{c3}\alpha l$ are comparable, respectively (see Supplemental Material [29]). It indicates that our evaluations of $J_{c2}$ and $J_{c3}$ were performed precisely. As shown in Fig. 8, at a glance, the field range of the dips seems to be correlated to the widths of the return branch since these values are almost the same. However, the temperature dependence of the characteristics field of the dip is weaker compared with the width of the return branch. The smaller temperature dependence suggests that the origin of the dip cannot be explained by the self-field effect. The matching effect may play a key role for the emergence of the

dip structure. It is also found that both $J_{c2}$ and $J_{c3}$ are larger than $J_c$ for the $H//c$ axis since planar defects parallel to the $ab$ plane do not work as vortex pinning centers for $J_c$ for the $H//c$ axis. Needless to say, the very large $J_c$ of CaKFe$_4$As$_4$ due to naturally-introduced defects is advantageous for applications. $J_c$ can be further enhanced by introducing defects by particle irradiation. Actually, it has been demonstrated that $J_c$ in the 122 system is enhanced significantly by swift particle irradiation [6]. Similar enhancements of $J_c$ in CaKFe$_4$As$_4$ by 3 MeV proton and 800 MeV Xe irradiation have been confirmed, and details of these effects will be published in a presented elsewhere.

Finally, we comment on a recent report by Singh *et al.* on the same material [40]. Although the qualitative behavior of $J_c$ in their report is similar to ours, the values of $J_c$ evaluated by them are significantly larger than ours.

**Conclusion**

In summary, CaKFe$_4$As$_4$ single crystals were synthesized and their critical current density was characterized including its anisotropy. The sharp onset of diamagnetism with low residual resistivity demonstrates their high phase purity, and MO images confirmed uniform flow of the shielding current. When the magnetic field is applied along the $c$ axis, the temperature dependence of $J_c$ shows non-monotonic temperature dependence. We interpret this anomalous behavior as an enhanced fish-tail effect rather than field-induced pinning centers. From the magnetization measurements and their analyses for a magnetic field parallel to the $ab$ plane, we have successfully evaluated two components of $J_c$, i.e., one parallel to the $ab$ plane ($J_{c2}$) and another parallel to the $c$ axis ($J_{c3}$). $J_{c2}$ turns out to be much larger than $J_{c3}$ and even larger that in-plane $J_c$ for the magnetic field parallel to the $c$ axis. TEM observations clarified the presence of planar defects along the $ab$ plane. We interpret the strong enhancement of $J_{c2}$ and $J_{c3}$ at low fields due to the geometrical matching of vortices to these planar defects.

**Acknowledgment**

This work was partially supported by a Grant-in-Aid for Scientific Research (A) (Grant No. 17H01141) by the Japan Society for the Promotion of Science (JSPS).

**Reference**

[1] Y. Kamihara, T. Watanabe, M. Hirano, and H. Hosono, Iron-based layered superconductor La[O$_{1-x}$F$_x$]FeAs ($x$ = 0.05-0.12) with $T_c$ = 26 K, J. Am. Chem. Soc. **130**, 3296 (2008).

[2] M. Rotter, M. Tegel, and D. Johrendt, Superconductivity at 38 K in the iron arsenide Ba$_{1-x}$K$_x$Fe$_2$As$_2$, Phys. Rev. Lett. **101,** 107006 (2008).

[3] H. Q. Yuan, J. Singleton, F. F. Balakirev, S. A. Baily, G. F. Chen, J. L. Luo, and N. L. Wang, Nearly isotropic superconductivity in (Ba,K)Fe$_2$As$_2$, Nature **457**, 565 (2009).

[4] M. M. Altarawneh, K. Collar, C. H. Mielke, N. Ni, S. L. Bud'ko, and P. C. Canfield, Determination of anisotropic $H_{c2}$ up to 60 T in $Ba_{0.55}K_{0.45}Fe_2As_2$ single crystals, Phys. Rev. B **78**, 220505 (2008).

[5] Y. Nakajima, Y. Tsuchiya, T. Taen, T. Tamegai, S. Okayasu, and M. Sasase, Enhancement of critical current density in Co-doped $BaFe_2As_2$ with columnar defects introduced by heavy-ion irradiation, Phys. Rev. B **80,** 012510 (2009).

[6] F. Ohtake, T. Taen, S. Pyon, T. Tamegai, S. Okayasu, T. Kambara, and H. Kitamura, Effects of heavy-ion irradiations in K-doped $BaFe_2As_2$, Physica C **518**, 47 (2015).

[7] T. Tamegai, T. Taen, H. Yagyuda, Y. Tsuchiya, S. Mohan, T. Taniguchi, Y. Nakajima, S. Okayasu, M. Sasase, H. Kitamura, T. Murakami, T. Kambara, and Y. Kanai, Effects of particle irradiations on vortex states in iron-based superconductors, Supercond. Sci. Technol. **25**, 084008 (2012).

[8] T. Taen, F. Ohtake, S. Pyon, T. Tamegai, and H. Kitamura, Critical current density and vortex dynamics in pristine and proton-irradiated $Ba_{0.6}K_{0.4}Fe_2As_2$, Supercond. Sci. Technol. **28**, 085003 (2015).

[9] L. Fang, Y. Jia, C. Chaparro, G. Sheet, H. Claus, M. A. Kirk, A. E. Koshelev, U. Welp, G. W. Crabtree, W. K. Kwok, S. Zhu, H. F. Hu, J. M. Zuo, H.-H. Wen, and B. Shen, High, magnetic field independent critical currents in $(Ba,K)Fe_2As_2$ crystals, Appl. Phys. Lett. **101**, 012601 (2012).

[10] M. Eisterer, Radiation effects on iron-based superconductors, Supercond. Sci. Technol. **31**, 013001 (2018).

[11] A. Iyo, K. Kawashima, T. Kinjo, T. Nishio, S. Ishida, H. Fujihisa, Y. Gotoh, K. Kihou, H. Eisaki, and Y. Yoshida, New-structure-type Fe-based superconductors: $CaAFe_4As_4$ ($A$ = K, Rb, Cs) and $SrAFe_4As_4$, ($A$ = Rb, Cs) J. Am. Chem. Soc. **138**, 1410 (2016).

[12] W. R. Meier, T. Kong, U. S. Kaluarachchi, V. Taufour, N. H. Jo, G. Drachuck, A. E. Böhmer, S. M. Saunders, A. Sapkota, A. Kreyssig, M. A. Tanatar, R. Prozorov, A. I. Goldman, F. F. Balakirev, A. Gurevich, S. L. Bud'ko, and P. C. Canfield, Anisotropic thermodynamic and transport properties of single-crystalline $CaKFe_4As_4$ Phys. Rev. B **94**, 064501 (2016).

[13] D. Mou, T. Kong, W. R. Meier, F. Lochner, L.-L. Wang, Q. Lin, Y. Wu, S. L. Bud'ko, I. Eremin, D. D. Johnson, P. C. Canfield, and Adam Kaminski, Enhancement of the superconducting gap by nesting in $CaKFe_4As_4$: A new high temperature superconductor, Phys. Rev. Lett. **117**, 277001 (2016).

[14] J. Cui, Q.-P. Ding, W. R. Meier, A. E. Bohmer, T. Kong, V. Borisov, Y. Lee, S. L. Bud'ko, R. Valenti, P. C. Canfield, and Y. Furukawa, Magnetic fluctuations and superconducting properties of $CaKFe_4As_4$ studied by $^{75}As$ NMR, Phys. Rev. B **96**, 104512 (2017).

[15] K. Cho, A. Fente, S. Teknowijoyo, M. A. Tanatar, K. R. Joshi, N. M. Nusran, T. Kong, W. R. Meier, U. Kaluarachchi, I. Guillamon, H. Suderow, S. L. Bud'ko, P. C. Canfield, and R. Prozorov,

Nodeless multiband superconductivity in stoichiometric single-crystalline CaKFe$_4$As$_4$, Phys. Rev. B **95**, 100502(R) (2017).

[16] R. Khasanov, W. R. Meier, Y. Wu, D. Mou, S. L. Bud'ko, I. Eremin, H. Luetkens, A. Kaminski, P. C. Canfield, and Alex Amato, In-plane magnetic penetration depth of superconducting CaKFe$_4$As$_4$, Phys. Rev. B **97**, 140503(R) (2018).

[17] U. S. Kaluarachchi, V. Taufour, A. Sapkota, V. Borisov, T. Kong, W. R. Meier, K. Kothapalli, B. G. Ueland, A. Kreyssig, R. Valenti, R. J. McQueeney, A. I. Goldman, S. L. Bud'ko, and P. C. Canfield, Pressure-induced half-collapsed-tetragonal phase in CaKFe$_4$As$_4$, Phys. Rev. B **96**, 140501 (2017).

[18] V. Borisov, P. C. Canfield, and R. Valenti, Trends in pressure-induced layer-selective half-collapsed tetragonal phases in the iron-based superconductor family *AeA*Fe$_4$As$_4$, Phys. Rev. B **98**, 064104 (2018).

[19] Y. Sun, Y. Tsuchiya, S. Pyon, T. Tamegai, C. Zhang, T. Ozaki, and Q. Li, Magneto-optical characterizations of FeTe$_{0.5}$Se$_{0.5}$ thin films with critical current density over 1 MA cm$^{-2}$, Supercond. Sci. Technol. **28**, 015010 (2015).

[20] C. P. Bean, Magnetization of high-field superconductors, Rev. Mod. Phys. **36**, 31 (1964).

[21] E. M. Gyorgy, R. B. van Dover, K. A. Jackson, L. F. Schneemeyer, and J. V. Waszczak, Anisotropic critical currents in Ba$_2$YCu$_3$O$_7$ analyzed using an extended Bean model, Appl. Phys. Lett. **55**, 283 (1989).

[22] U. Welp, W. K. Kwok, G. W. Crabtree, K. G. Vandervoort, and J. Z. Liu, Magnetization hysteresis and flux pinning in twinned and untwinned YBa$_2$Cu$_3$O$_{7-x}$ single crystals, Appl. Phys. Lett. **57**, 84 (1990).

[23] M. Daeumling, J. M. Seuntjens, and D. C Larbalestier, Oxygen-defect flux pinning, anomalous magnetization and intra-grain granularity in YBa$_2$Cu$_3$O$_{7-\delta}$, Nature **346**, 332 (1990).

[24] N. Chikumoto, M. Konczykowski, N. Motohira, and A. P. Malozemoff, Flux-creep crossover and relaxation over surface barriers in Bi$_2$Sr$_2$CaCu$_2$O$_8$ Crystals, Phys. Rev. Lett. **69**, 1260 (1992).

[25] T. Kimura, K. Kishio, T. Kobayashi, Y. Nakayama, N. Motohira, K. Kitazawa, and K. Yamafuji、Compositional dependence of transport anisotropy in large (La,Sr)$_2$CuO$_4$ single crystals and second peak in magnetization curves, Physica C **192**, 247 (1992).

[26] S. Ishida, D. Song, H. Ogino, A. Iyo, and H. Eisaki, Doping-dependent critical current properties in K, Co, and P-doped BaFe$_2$As$_2$ single crystals, Phys. Rev. B **95**, 014517 (2017).

[27] H. Yang, H. Luo, Z. Wang, and H. H. Wen, Fishtail effect and the vortex phase diagram of single crystal Ba$_{0.6}$K$_{0.4}$Fe$_2$As$_2$, Appl. Phys. Lett. **93**, 142506 (2008).

[28] T. Taen, Y. Tsuchiya, Y. Nakajima, and T. Tamegai, Superconductivity at $T_c$ ~ 14 K in single-crystalline FeTe$_{0.61}$Se$_{0.39}$, Phys. Rev. B **80**, 092502 (2009).

[29] See Supplemental Material for the details of the estimation of anisotropic critical current density

from magnetization measurements.

[30] G. P. Mikitik and E. H. Brandt, Critical state in thin anisotropic superconductors of arbitrary shape, Phys. Rev. B **62**, 6800 (2000).

[31] N. Ito, S. Pyon, T. Kambara, A. Yoshida, S. Okayasu, A. Ichinose, and T. Tamegai, Anisotropy of critical current densities in $Ba_{1-x}K_xFe_2As_2$ and $Ba(Fe_{1-x}Co_x)_2As_2$ with splayed columnar defects, IOP Conf. Ser.: J. Phys.: Conf. Ser. **1054**, 012020 (2018).

[32] W. R. Meier, T. Kong, S. L. Bud'ko, and P. C. Canfield, Optimization of the crystal growth of the superconductor $CaKFe_4As_4$ from solution in the $FeAs$-$CaFe_2As_2$-$KFe_2As_2$ system, Phys. Rev. Mater. **1** 013401 (2017).

[33] R. Prozorov, N. Ni, M. A. Tanatar, V. G. Kogan, R. T. Gordon, C. Martin, E. C. Blomberg, P. Prommapan, J. Q. Yan, S. L. Bud'ko, and P. C. Canfield, Vortex phase diagram of $Ba(Fe_{0.93}Co_{0.07})_2As_2$ single crystals, Phys. Rev. B **78**, 224506 (2008).

[34] Y. Sun, T. Taen, Y. Tsuchiya, Q. Ding, S. Pyon, Z. Shi, and T. Tamegai, Large, Homogeneous, and Isotropic Critical Current Density in Oxygen-Annealed $Fe_{1+y}Te_{0.6}Se_{0.4}$ Single Crystal, Appl. Phys. Express **6**, 043101 (2013).

[35] M. Sato, T. Shibauchi, S. Ooi, T. Tamegai, and M. Konczykowski, Recoupling of decoupled vortex liquid by columnar defects in $Bi_2Sr_2CaCu_2O_{8+y}$, Phys. Rev. Lett. **79**, 3759 (1997).

[36] K. Itaka, T. Shibauchi, M. Yasugaki, T. Tamegai, and S. Okayasu, Asymmetric field profile in Bose glass phase of irradiated $YBa_2Cu_3O_{7-\delta}$: loss of interlayer coherence around 1/3 of matching field, Phys. Rev. Lett. **86**, 5144 (2001).

[37] A. Park, S. Pyon, K. Ohara, N. Ito, T. Tamegai, T. Kambara, A. Yoshida, and A. Ichinose, Field-driven transition in the $Ba_{1-x}K_xFe_2As_2$ superconductor with splayed columnar defects, Phys. Rev. B **97**, 064516 (2018).

[38] A. N. Lykov, Pinning in superconducting films with triangular lattice of holes, Solid State Commun. **86**, 531 (1993).

[39] J. I. Martin, M. Velez, A. Hoffmann, I. K. Schuller, and J. L. Vicent, Temperature dependence and mechanisms of vortex pinning by periodic arrays of Ni dots in Nb films, Phys. Rev. B **62**, 9110 (2000).

[40] S. J. Singh, M. Bristow, W. R. Meier, P. Taylor, S. J. Blundell, P. C. Canfield, and A. I. Coldea1, Ultrahigh critical current densities, the vortex phase diagram, and the effect of granularity of the stoichiometric high-$T_c$ superconductor $CaKFe_4As_4$, Phys. Rev. Mater. **2**, 074802 (2018).

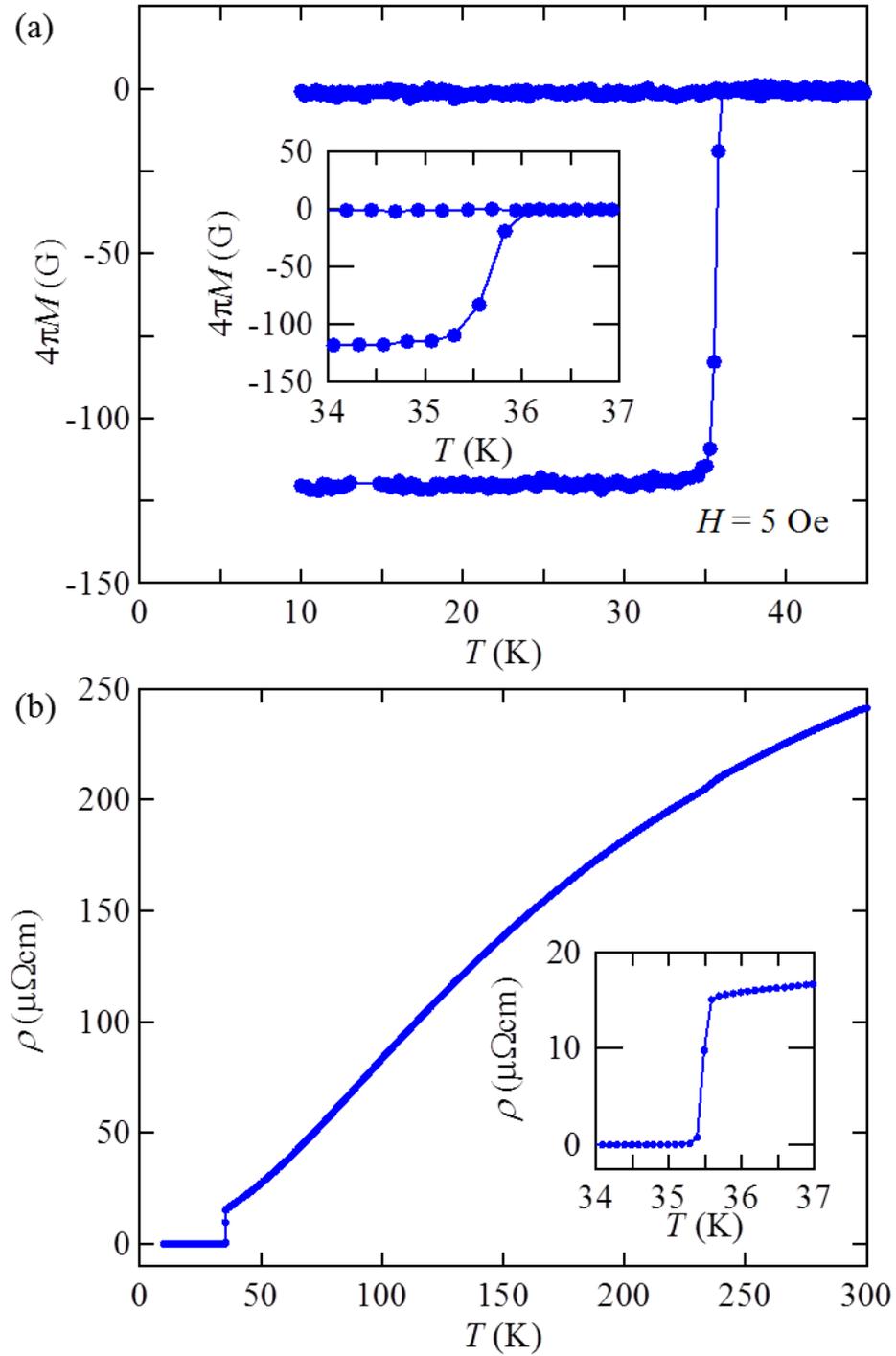

Fig. 1. (a) Temperature dependence of magnetization ($M$) at 5 Oe for CaKFe$_4$As$_4$ single crystal. The inset shows a blow-up of $M$-$T$ near $T_c$. Dimensions of the sample are 0.138 x 0.083 x 0.0040 cm$^3$. (b) Temperature dependence of electrical resistivity ($\rho$) for CaKFe$_4$As$_4$ single crystal. The inset shows a blow-up of $\rho$-$T$ near $T_c$.

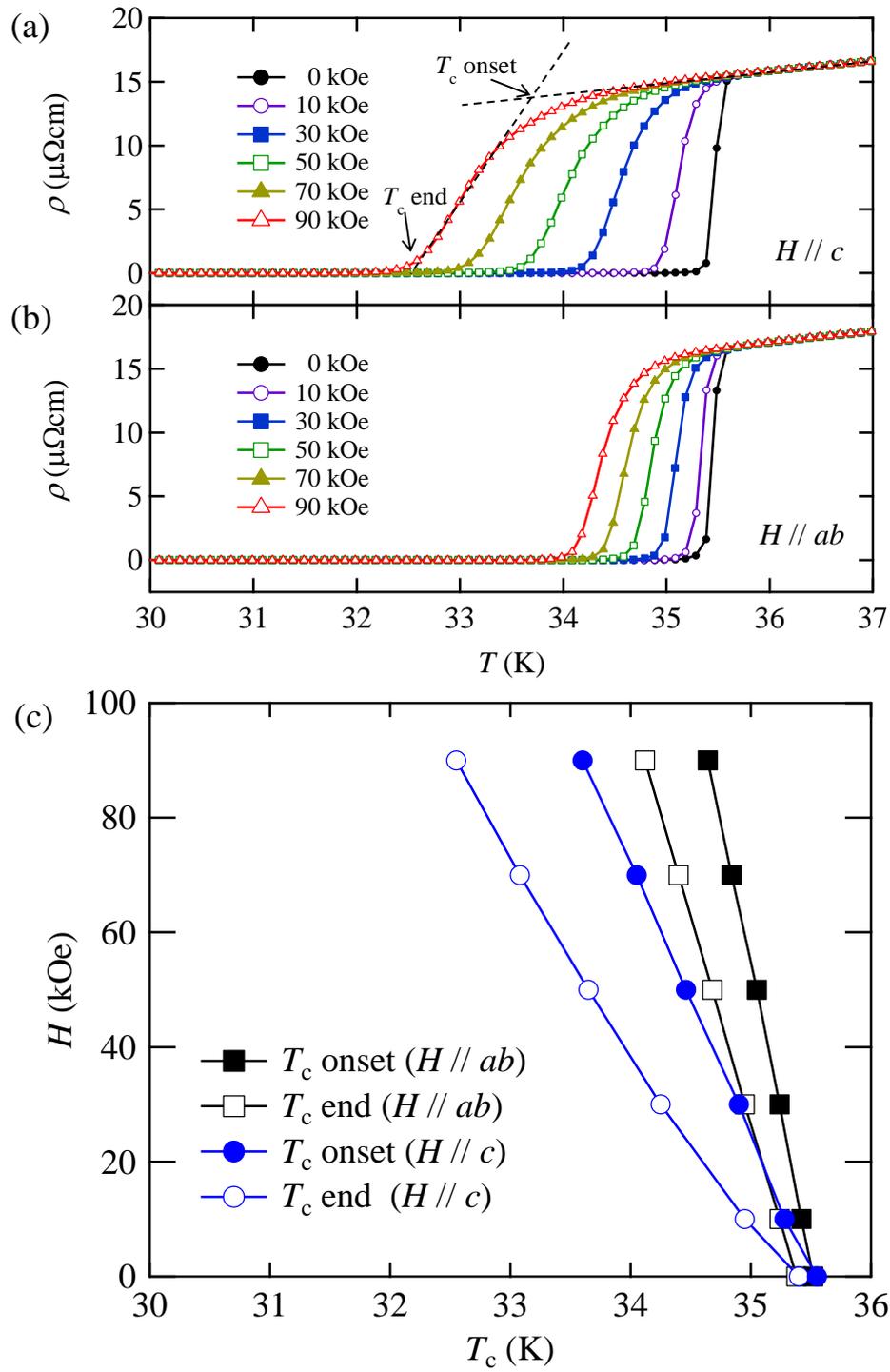

Fig. 2. Temperature dependence of electrical resistivity in CaKFe$_4$As$_4$ single crystal under various magnetic fields parallel to (a) $c$ axis and (b) $ab$ plane. (c) Anisotropic $H_{c2}$ evaluated from temperature-dependent resistivity presented in (a) and (b).

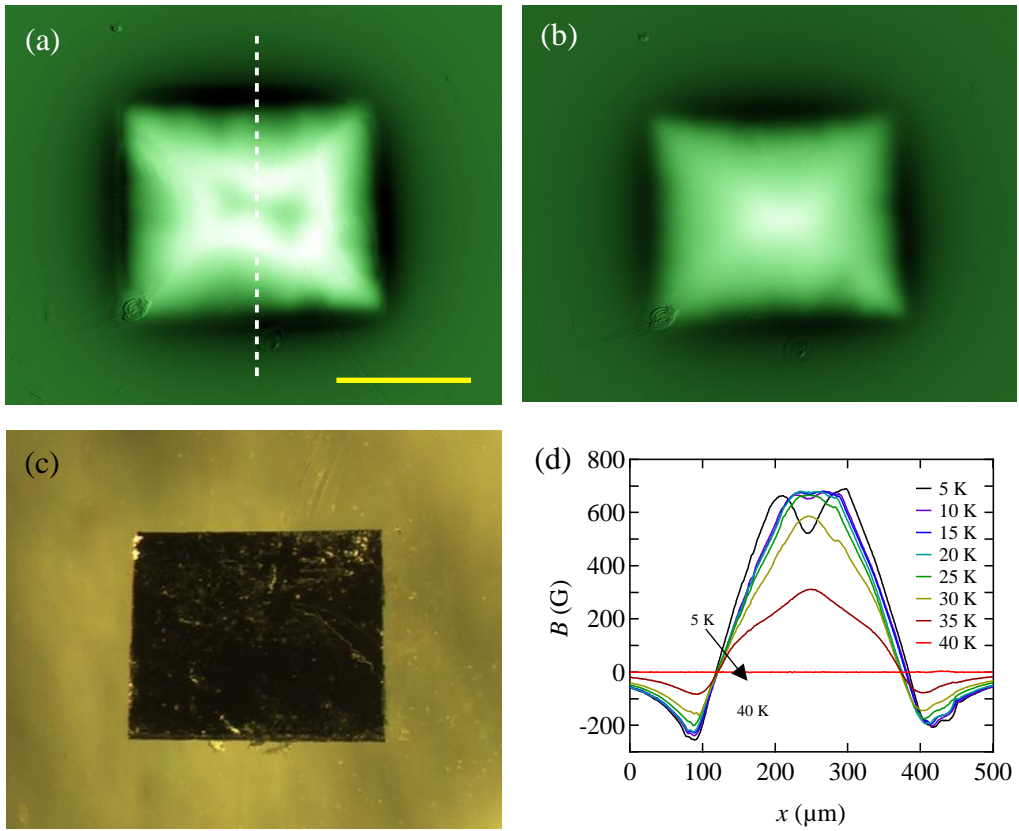

Fig. 3. Differential MO images of CaKFe$_4$As$_4$ single crystal in the remanent state at (a) 5 K and (b) 35 K after cycling the field up to 1.6 kOe for 0.2 s. Dimensions of the sample are 0.036 x 0.030 x 0.0005 cm$^3$. The spatial resolution of the MO image is approximately 1 μm. (c) Optical micrograph of CaKFe$_4$As$_4$ single crystal. (d) Local magnetic induction profiles at different temperatures taken along the dotted line in (a). The yellow bar in (a) corresponds to 200 μm.

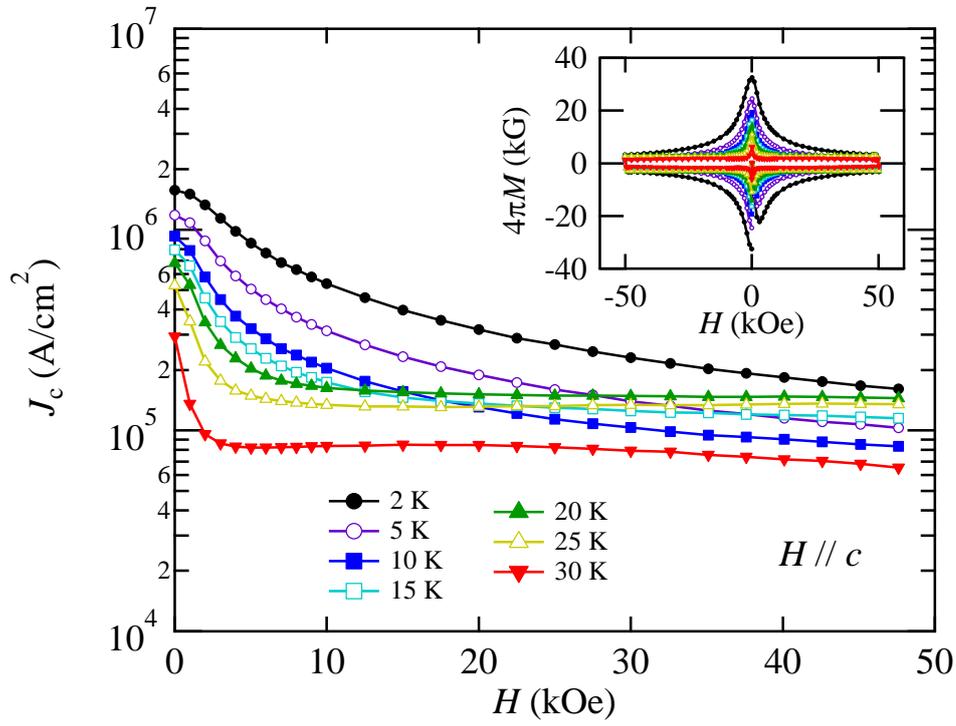

Fig. 4. Magnetic field dependence of magnetic $J_c$ in CaKFe$_4$As$_4$ single crystal at various temperatures for the field parallel to the $c$ axis. The same sample was used for Fig. 1(a) and this figure. The inset shows the magnetic field dependence of the magnetization in CaKFe$_4$As$_4$ single crystal at various temperatures.

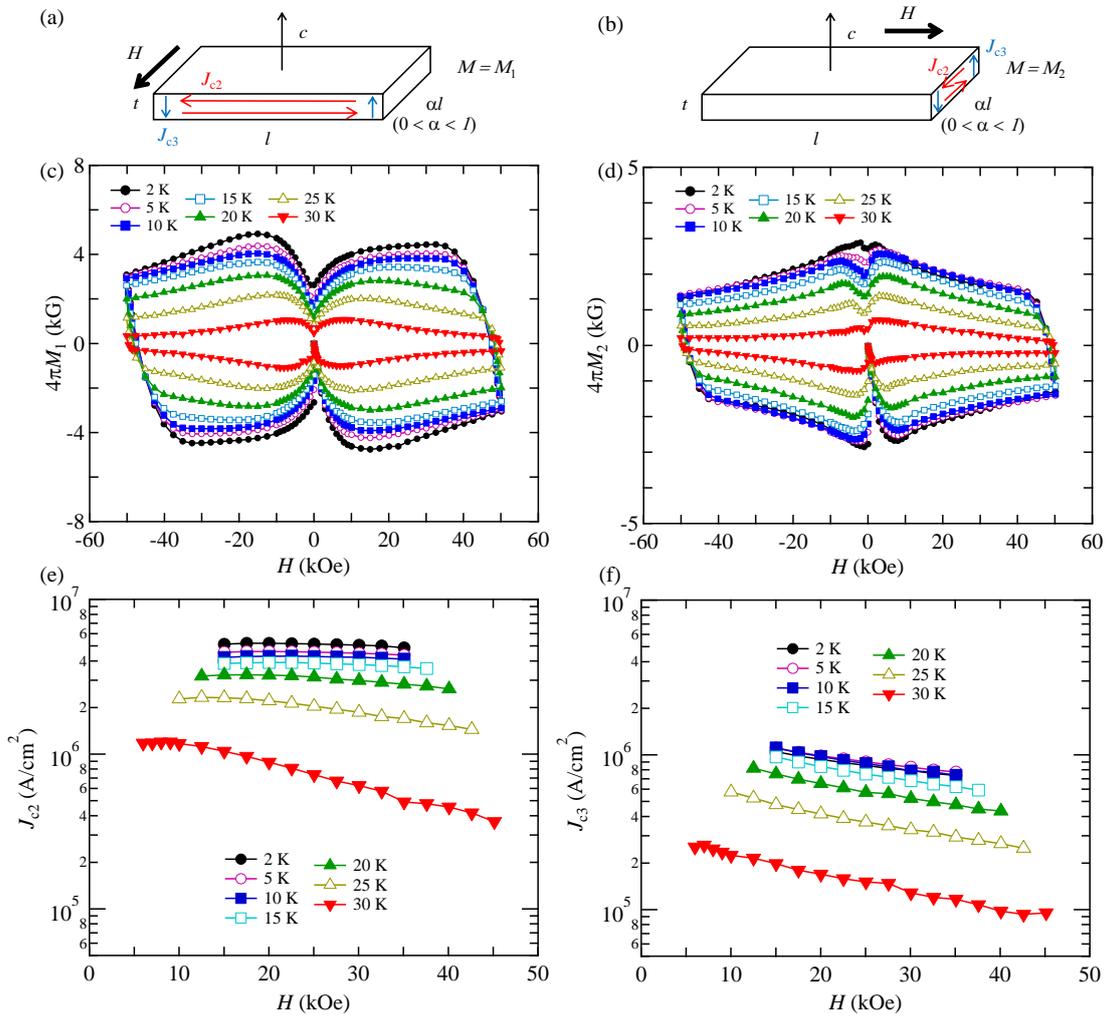

Fig. 5. Two components of critical current density $J_{c2}$ and $J_{c3}$ for fields parallel to the $ab$ plane along (a) short and (b) long edges of the crystal. Dimension of the sample are 0.046 x 0.0094 x 0.0029 cm$^3$. (c), (d) The magnetic field dependences of in-plane magnetization for the cases of (a) $M_1$ and (b) $M_2$, respectively. (e), (f) The magnetic field dependences of $J_{c2}$ and $J_{c3}$ calculated from the analyses described in the Supplemental Material [29].

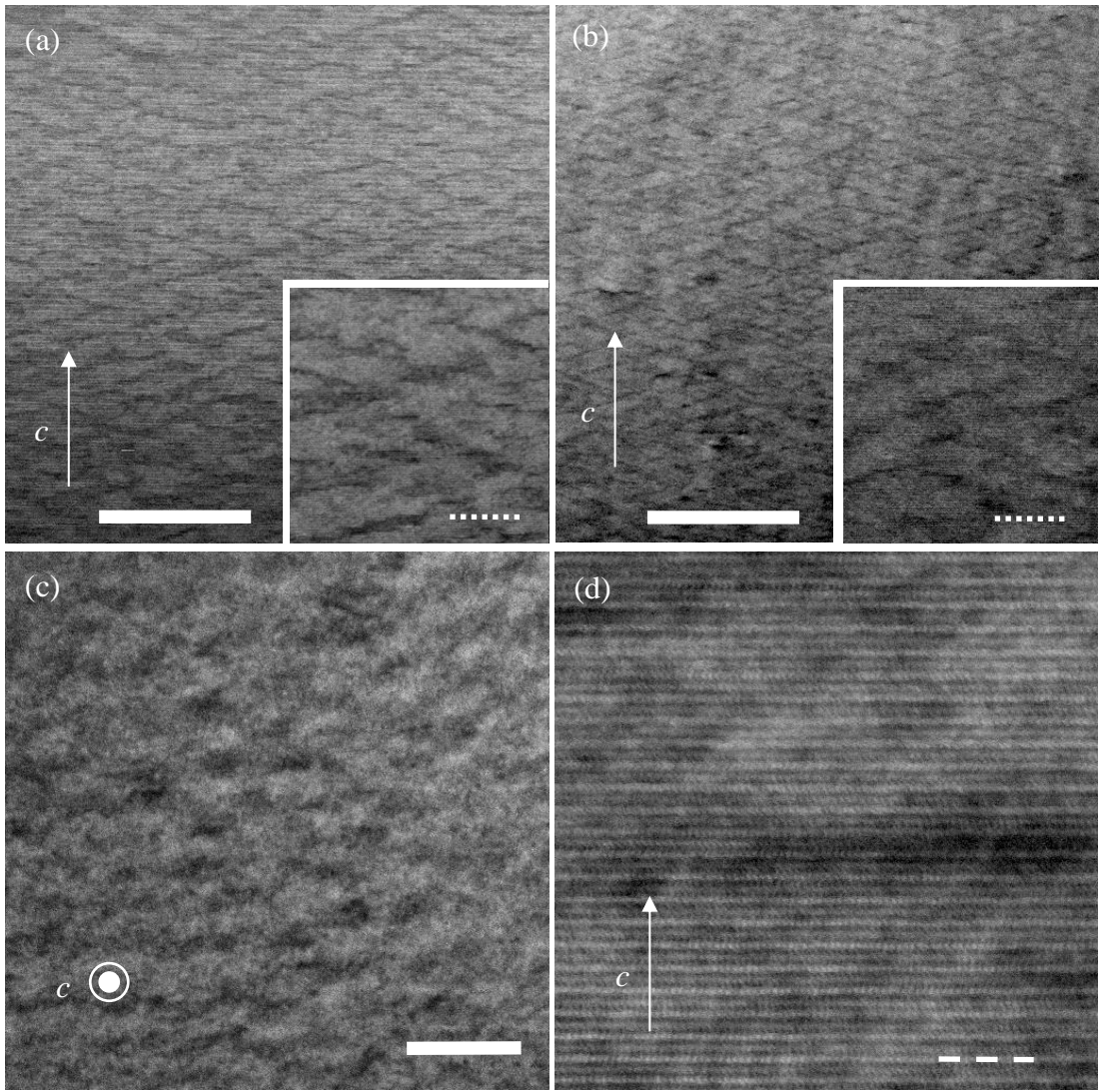

Fig. 6. STEM images of CaKFe$_4$As$_4$ for an electron beam injected along the (a) *a* axis, (b) *b* axis, and (c) *c* axis. Insets of (a) and (b) are the higher magnification image for the *a*-axis and *b*-axis injection, respectively. (d) Higher-resolution STEM image of CaKFe$_4$As$_4$ for an electron beam injected along the *a* axis. The solid lines in (a)-(c), dashed line in (d), and dotted line in the inset correspond to 200, 5, and 50 nm, respectively.

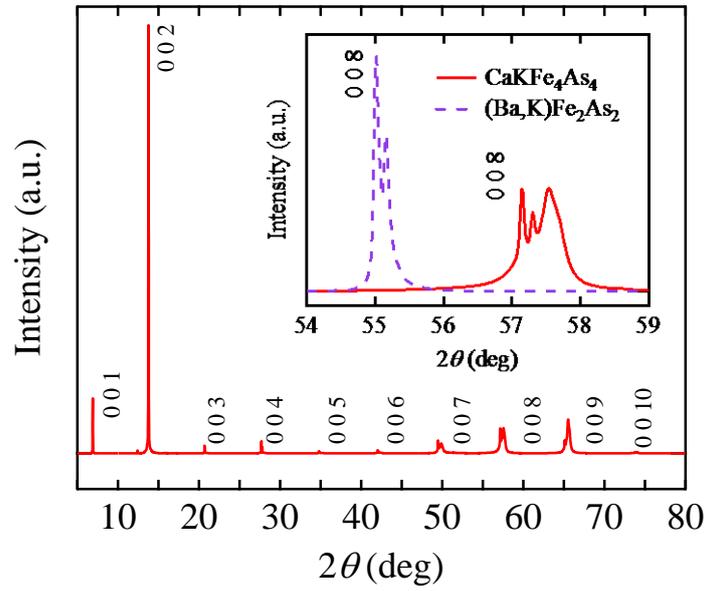

Fig. 7. Single-crystal x-ray diffraction pattern of $CaKFe_4As_4$. The inset shows a comparison of the (0 0 8) peaks for the $CaKFe_4As_4$ and $(Ba,K)Fe_2As_2$ single crystals.

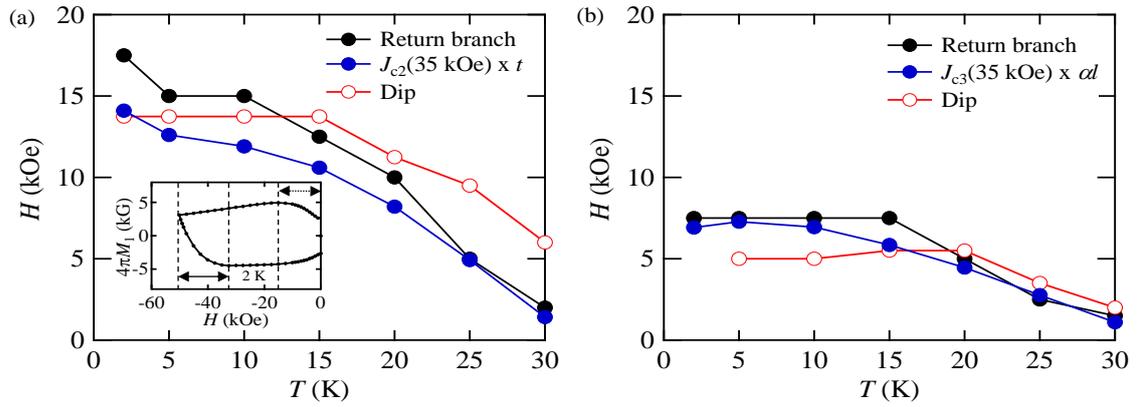

Fig. 8. (a), (b) Temperature dependence of the field ranges of the return branches and dips, and products of $J_{c2}$ and $t$ and that of $J_{c3}$ and $\alpha l$, referred from the magnetization hysteresis loops in Figs. 5(c) and 5(d), respectively. The inset shows an example of the hysteresis loop, where an arrow and dotted arrow indicate the ranges of the return branch and dip, respectively. An arrow or dotted arrow indicates the range of a return branch or a dip, respectively.

# Estimation of anisotropic critical current density $J_c$ from magnetization measurements

## 1 Detail of analyses

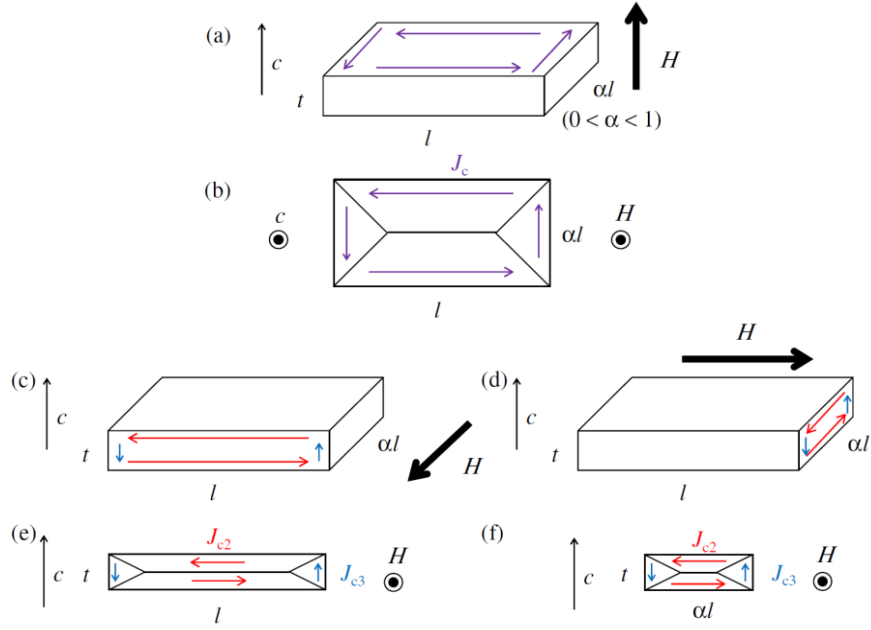

Figure S1 : Rectangular samples in applying magnetic field (a) parallel to the $c$-axis and (b), (c) perpendicular to $c$-axis. Critical currents densities of $J_{c2}$ and $J_{c3}$, corresponding $d$-lines are also described in each lower figures.

Here, we consider a critical state in a thin platelet superconductor (thickness $t$) with a layered structure and tetragonal symmetry with its symmetry axis along the $c$-axis. Without loss of generality, we assume that the length ($l$) of the sample is larger than the width ($w = \alpha l, 0 < \alpha < 1$). When the magnetic field is applied parallel to the $c$-axis, tetragonal symmetry imposes isotropic in-plane critical current density, $J_c$, resulting in the current discontinuity line as shown in figure S1(b). $J_c$ can be evaluated from $\Delta M$, which is the width of $M - H$ hysteresis loop, using the extended Bean model;

$$J_c = \frac{20\Delta M}{\alpha l(1 - \frac{\alpha}{3})} \qquad (1)$$

On the other hand, when the magnetic field is applied parallel to the $ab$-plane, we have two independent critical current densities, one along the $ab$-pane, $J_{c2}$, and another along the $c$-axis, $J_{c3}$ as shown in figures S1(c) and (d). These two $J_c$ generate $d$-lines on the lateral surface of the sample as shown in figures

S1(e) and (f). In order to evaluate these two components of $J_c$, we need two independent magnetization measurements for fields along the short ($M_1$) and long ($M_2$) edges of the sample.

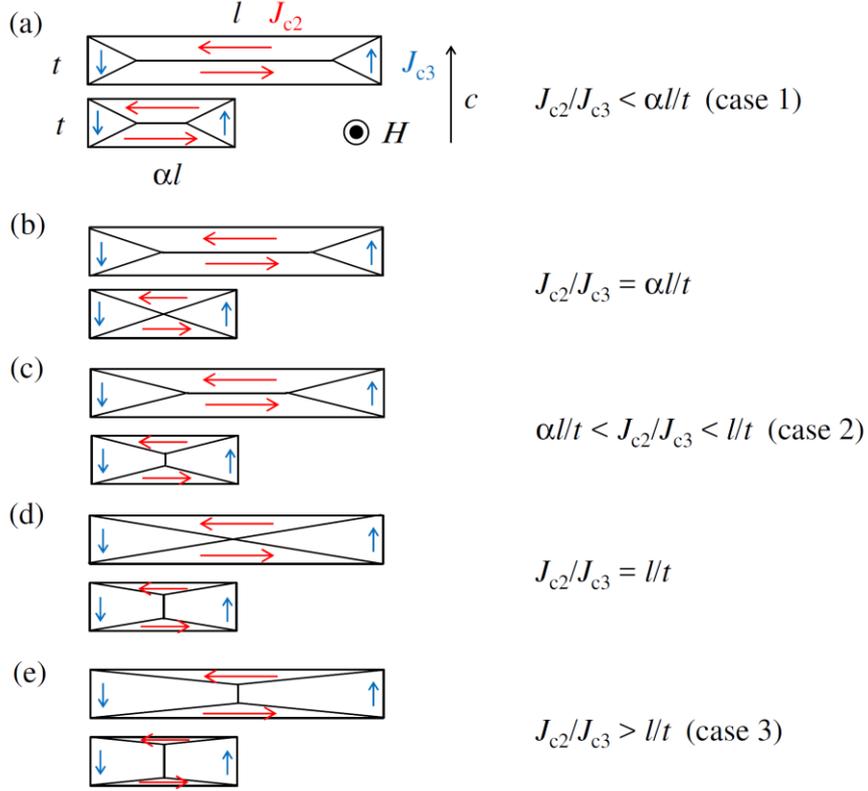

Figure S2 : (a)-(e) Schematic drawings of $d$ lines and corresponding critical current densities of $J_{c2}$ and $J_{c3}$. In right parts, conditions of $\frac{J_{c2}}{J_{c3}}$ are shown.

Figures S2(a)-(e) show several cases of $d$-lines on two orthogonal lateral surfaces when the magnetic field is applied parallel to the $ab$-plane. Numbers and locations of $d$-lines depend both on the ratios of the lengths of edges on the lateral surface, $\alpha \frac{l}{t}$, $\frac{l}{t}$, and the ratio of $J_{c2}$ and $J_{c3}$. Conditions for the all $d$-line patterns are summarized in Figures S2(a)-(e). There are three cases; $0 < \frac{J_{c2}}{J_{c3}} < \alpha \frac{l}{t}$ (case 1), $\alpha \frac{l}{t} < \frac{J_{c2}}{J_{c3}} < \frac{l}{t}$ (case 2) and $\frac{J_{c2}}{J_{c3}} > \frac{l}{t}$ (case 3). $\Delta M_1$ and $\Delta M_2$ for the three cases are expressed by the following equations (2) and (3).

$$\Delta M_1 = \begin{cases} \dfrac{tJ_{c2}}{20}\left(1 - \dfrac{tJ_{c2}}{3lJ_{c3}}\right) & \left(0 < \dfrac{J_{c2}}{J_{c3}} < \dfrac{l}{t}; \text{case 1 and 2}\right) \\ \dfrac{lJ_{c3}}{20}\left(1 - \dfrac{lJ_{c3}}{3tJ_{c2}}\right) & \left(\dfrac{J_{c2}}{J_{c3}} > \dfrac{l}{t}; \text{case 3}\right) \end{cases} \quad (2)$$

$$\Delta M_2 = \begin{cases} \dfrac{tJ_{c2}}{20}\left(1 - \dfrac{tJ_{c2}}{3\alpha lJ_{c3}}\right) & \left(0 < \dfrac{J_{c2}}{J_{c3}} < \alpha \dfrac{l}{t}; \text{case 1}\right) \\ \dfrac{\alpha lJ_{c3}}{20}\left(1 - \dfrac{\alpha lJ_{c3}}{3tJ_{c2}}\right) & \left(\dfrac{J_{c2}}{J_{c3}} > \alpha \dfrac{l}{t}; \text{case 2 and 3}\right) \end{cases} \quad (3)$$

Using $\Delta M_1$ and $\Delta M_2$ described in equations (2) and (3), $J_{c2}$ and $J_{c3}$ can be calculated for each case. For cases 1 and 3, $J_{c2}$ and $J_{c3}$ can be expressed analytically as shown in equations (4), (5), (8) and (9). For case 2, however, cubic equations shown in equations (6) and (7) should be solved to obtain $J_{c2}$ and $J_{c3}$.

Case 1 : $\left(0 < \dfrac{J_{c2}}{J_{c3}} < \alpha\dfrac{l}{t}\right)$

$$J_{c2} = \frac{20(\alpha\Delta M_2 - \Delta M_1)}{t(\alpha - 1)} \tag{4}$$

$$J_{c3} = \frac{20(\alpha\Delta M_2 - \Delta M_1)^2}{3\alpha l(\alpha - 1)(\Delta M_2 - \Delta M_1)} \tag{5}$$

Case 2 : $\left(\alpha\dfrac{l}{t} < \dfrac{J_{c2}}{J_{c3}} < \dfrac{l}{t}\right)$

$$\left(\frac{\alpha t^3}{3} - \frac{\alpha^2 t^3}{27}\right) J_{c2}^3 - 20\left(\frac{\alpha t^2}{3}\Delta M_1 + t^2 \Delta M_2\right) J_{c2}^2 \\ + 800 t \Delta M_1 \Delta M_2 J_{c2} - 8000(\Delta M_1)^2 \Delta M_2 = 0 \tag{6}$$

$$\left(\frac{\alpha^3 l^3}{3} - \frac{\alpha^4 l^3}{27}\right) J_{c3}^3 - 20\left(\alpha^2 l^2 \Delta M_1 + \frac{\alpha^2 l^2}{3}\Delta M_2\right) J_{c3}^2 \\ + 800 \alpha l \Delta M_1 \Delta M_2 J_{c3} - 8000 \Delta M_1 (\Delta M_2)^2 = 0 \tag{7}$$

Case 3 : $\left(\dfrac{l}{t} < \dfrac{J_{c2}}{J_{c3}}\right)$

$$J_{c2} = \frac{20(\Delta M_2 - \alpha^2 \Delta M_1)^2}{3\alpha t(1 - \alpha)(\Delta M_2 - \alpha\Delta M_1)} \tag{8}$$

$$J_{c3} = \frac{20(\Delta M_2 - \alpha^2 \Delta M_1)}{\alpha l(1 - \alpha)} \tag{9}$$

When we evaluate $J_{c2}$ and $J_{c3}$, we first need to decide which case's formula should be used. This is determined from experimentally obtained $\Delta M_1$ and $\Delta M_2$ data. Here we define $\sigma = \dfrac{J_{c2}}{J_{c3}}$ and $\mu(\sigma) = \dfrac{\Delta M_1}{\Delta M_2}$. From equations (2) and (3), $m(\sigma)$ is calculated as shown in equation (10).

$$\mu(\sigma) = \begin{cases} \dfrac{1 - \frac{t}{3l}\sigma}{1 - \frac{t}{3\alpha l}\sigma} & \left(0 < \sigma < \alpha\dfrac{l}{t}; \text{case 1}\right) \\ \dfrac{t}{\alpha l}\sigma \dfrac{1 - \frac{t}{3l}\sigma}{1 - \frac{\alpha l}{3t}\frac{1}{\sigma}} & \left(\alpha\dfrac{l}{t} < \sigma < \dfrac{l}{t}; \text{case 2}\right) \\ \dfrac{1}{\alpha}\dfrac{1 - \frac{l}{3t}\frac{1}{\sigma}}{1 - \frac{\alpha l}{3t}\frac{1}{\sigma}} & \left(\sigma > \dfrac{l}{t}; \text{case 3}\right) \end{cases} \tag{10}$$

The derivatives of $\mu$, $\dfrac{d\mu(\sigma)}{d\sigma}$ is positive in all cases. So $\mu(\sigma)$ should monotonically increase as a function of $\sigma$. Values of $\mu(\sigma)$ for special values of $\sigma$ are summarized in equation (11).

$$\mu(\sigma) = \begin{cases} 1 & (\sigma \to 0) \\ \dfrac{3-\alpha}{2} & \left(\sigma = \alpha\dfrac{l}{t}\right) \\ \dfrac{2}{\alpha(3-\alpha)} & \left(\sigma = \dfrac{l}{t}\right) \\ \dfrac{1}{\alpha} & (\sigma \to \infty) \end{cases} \quad (11)$$

From these relations, three cases are redefined using $\mu$, as summarized in table 1.

Table S1: Three cases of $d$-line patterns described using $\sigma = \dfrac{J_{c2}}{J_{c3}}$ or $\mu = \dfrac{\Delta M_1}{\Delta M_2}$

| case | range of $\sigma$ | range of $\mu$ |
|------|-------------------|----------------|
| 1 | $0 < \sigma < \alpha\dfrac{l}{t}$ | $1 < \mu < \dfrac{3-\alpha}{2}$ |
| 2 | $\alpha\dfrac{l}{t} < \sigma < \dfrac{l}{t}$ | $\dfrac{3-\alpha}{2} < \mu < \dfrac{2}{\alpha(3-\alpha)}$ |
| 3 | $\sigma > \dfrac{l}{t}$ | $\dfrac{2}{\alpha(3-\alpha)} < \mu < \dfrac{1}{\alpha}$ |

It should be noted that sometimes the value of $\mu$ is out of the range for three cases. For example, near the self-field and return branch at high fields, $\mu$ may show anomalous values due to strong field dependence of $J_c$ or counter-flow of critical currents. When the sample is partially cleaved it is possible that $\mu$ is out range of the conditions. Furthermore, sample dimensions affect the range of $\mu$. If $\alpha$ is close to 1, ranges of $\mu$ and $\sigma$ are too narrow to analyze. In addition, if $\mu$ is close to $\alpha$ and $\dfrac{1}{\alpha}$, $J_{c2}$ and $J_{c3}$ diverge.

In summary, anisotropic $J_{c2}$ and $J_{c3}$ can be calculated using $\Delta M_1$ and $\Delta M_2$ from equations (4)-(9), which were evaluated from two independent magnetization measurements where magnetic fields are applied parallel to the $ab$-plane. Which case of formula should be used is decided by the value of $\mu(=\dfrac{\Delta M_1}{\Delta M_2})$. For proper evaluations of $J_{c2}$ and $J_{c3}$, the dimensions of sample size should be carefully chosen.

## 2 Experiments

The details how we evaluated $J_{c2}$ and $J_{c3}$ is explained below. The dimensions of rectangular sample were $0.0464 \times 0.0094 \times 0.0029$ cm$^3$. So $l$, $t$, and $\alpha$ are 0.0464, 0.0029, and 0.203, respectively. The values of $\dfrac{3-\alpha}{2}$, $\dfrac{2}{\alpha(3-\alpha)}$, and $\dfrac{1}{\alpha}$, are 1.399, 3.529, and 4.936, respectively. From two kinds of magnetization measurements, $\Delta M_1$ and $\Delta M_2$ were evaluated as shown in figures 5 (a) and (b) in the main text. Magnetic field dependences of $\dfrac{\Delta M_1}{\Delta M_2}$ at different temperatures are shown in figure S3. $\Delta M_1$ and $\Delta M_2$ are selected with reference to figure S3. When $\dfrac{\Delta M_1}{\Delta M_2}$ is less than unity, those data points were ignored. Evaluated $J_{c2}$ and $J_{c3}$ are summarized in figures S4 (a) and (b). In figure 5 (a) in the main text, reduction of magnetization around the self-field (-15 kOe $\sim$ 15 kOe) and return branch at high fields (above 35 kOe and below -35 kOe) are observed.

As described in the previous section, the extended Bean model with a field-independent $J_c$ cannot be adopted. So the data of $J_{c2}$ and $J_{c3}$ around self-field and return branch at high field are omitted in figures 5 (c) and (d) in the main text.

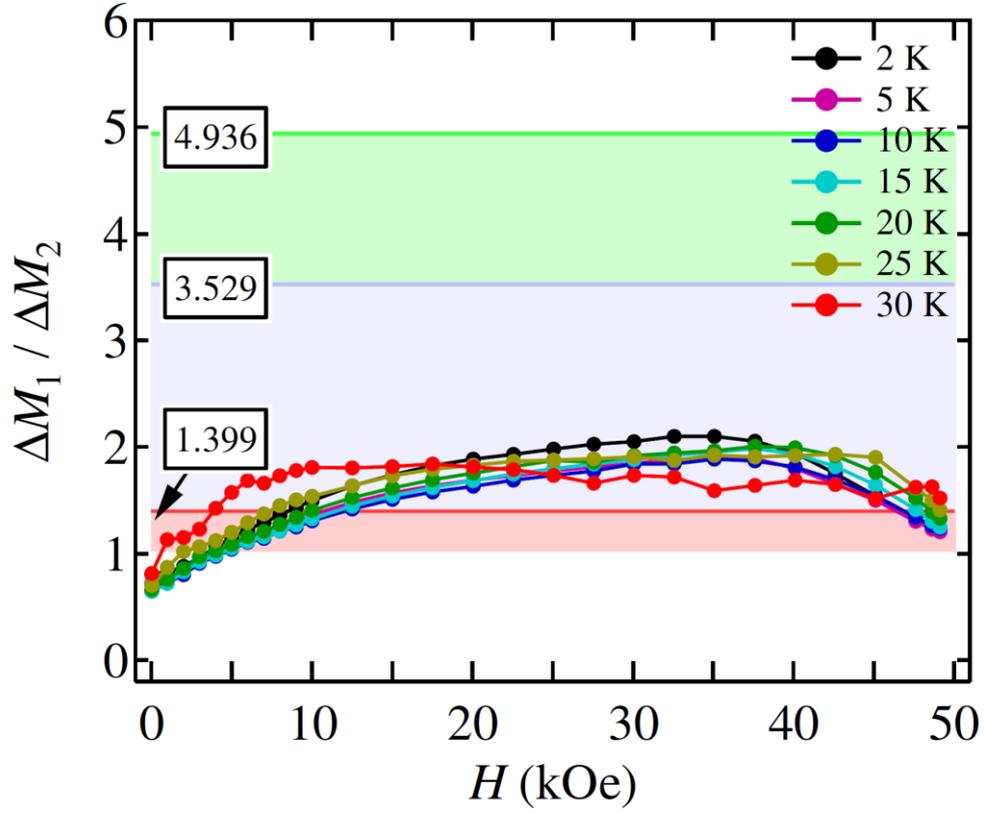

Figure S3 : Magnetic field dependence of $\frac{\Delta M_1}{\Delta M_2}$ at various temperatures. Red, blue, and green hatched areas indicate the case 1, 2, 3 respectively.

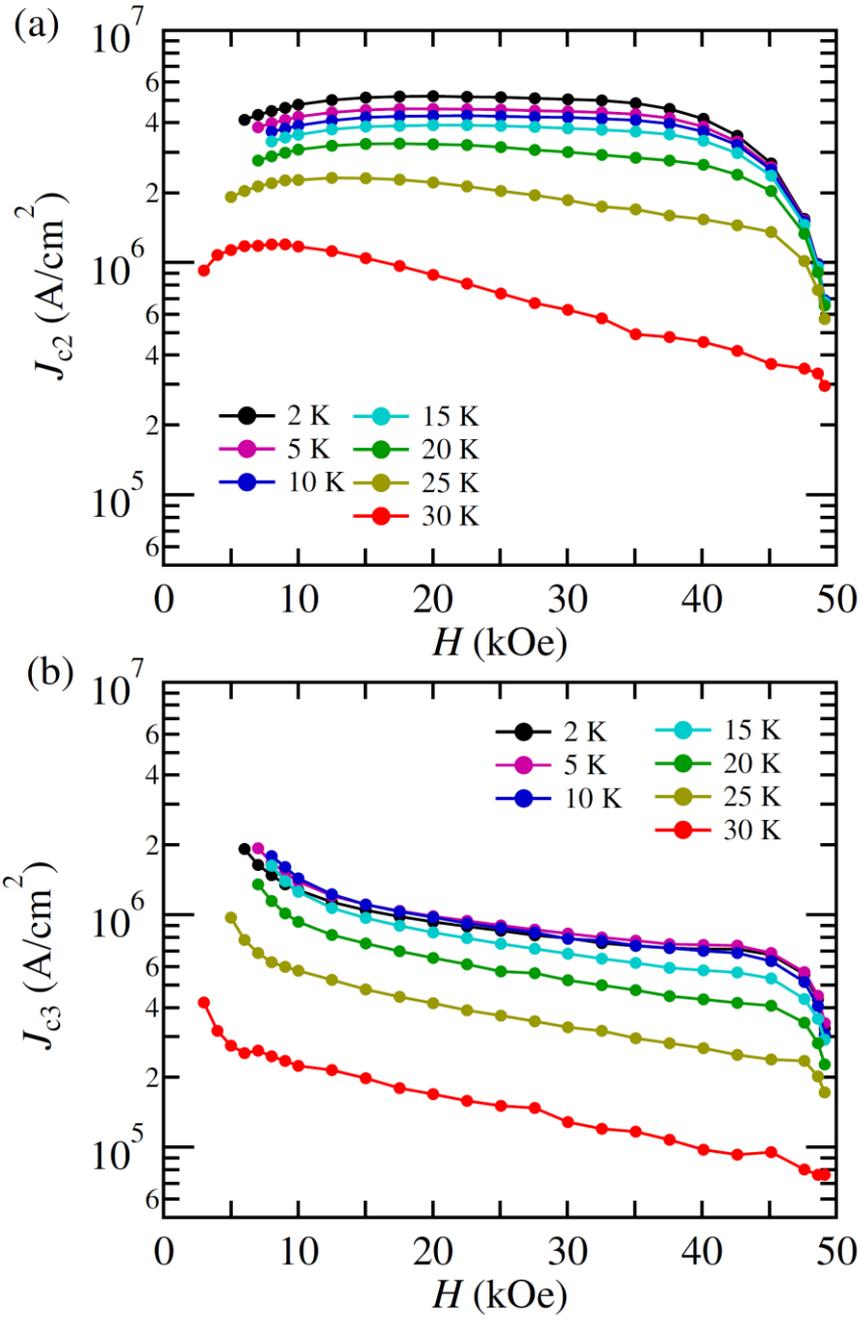

Figure S4 : Estimated magnetic field dependence of (a) $J_{c2}$ (b) $J_{c3}$ in CaKFe$_4$As$_4$ a single crystal at various temperatures.